\documentstyle[twoside,fleqn,espcrc2,epsfig]{article}
\newcommand{\be}{\begin{equation}}
\newcommand{\ee}{\end{equation}}
\newcommand{\ba}{\begin{eqnarray}}
\newcommand{\ea}{\end{eqnarray}}
\newcommand{\NL}{\nonumber \\ }
\newcommand{\ra}{\rightarrow}
\def\vep{\varepsilon}
\title{Chaos analyses in both phases of QED and QCD} 
\author{Tam\'as S. Bir\'o\address{Research Institute for 
Particle and Nuclear Physics, Pf.49, H-1525 Budapest, Hungary},
Natascha H\"ormann\address{Institute for Nuclear Physics, TU-Wien,
Wiedner Hauptstra\ss e 8-10, A-1040 Vienna, Austria},
Harald Markum$^{\rm b}$, and
Rainer Pullirsch$^{\rm b}$}
\begin{document}
\begin{abstract} 
 	We analyze the leading Lyapunov exponents of U(1) and
  	SU(2) gauge field configurations on the lattice
        which are initialized by quantum Monte Carlo simulations.
        We find that configurations in the strong coupling phase
        are substantially more chaotic than in deconfinement.
\end{abstract}
\maketitle
 
\section{Motivation}

The study of chaotic dynamics of classical field configurations in field
theory finds its motivation in phenomenological applications as well 
as for the understanding of basic principles. 
The role of chaotic field dynamics for the confinement of quarks is a 
longstanding question. Here, we analyze the leading Lyapunov exponents 
of compact U(1) and of SU(2)-Yang-Mills field configurations on the 
lattice. The real-time evolution of the classical field equations 
was initialized from Euclidean equilibrium configurations created 
by quantum Monte Carlo simulations. This way we expect to see
a coincidence between the strong coupling phase and the strength of
chaotic behavior in lattice simulations.

After reviewing essential
definitions of the physical quantities describing chaos
and their computation in lattice gauge theory \cite{BOOK}
we outline our method for the extraction of starting
configurations of a three dimensional Hamiltonian dynamics  from
four dimensional Euclidean field configurations.
Our results are then presented by showing an example of the
exponential divergence of small initial distances between
nearby field configurations. It is followed by a detailed study
of the maximal Lyapunov exponent and average plaquette energy
as a function of the coupling strength.

\section{Classical chaotic dynamics }

Chaotic dynamics in general is characterized by the
spectrum of Lyapunov exponents. These exponents, if they are positive,
reflect an exponential divergence of initially adjacent configurations.
In case of symmetries inherent in the Hamiltonian of the system
there are corresponding zero values of these exponents. Finally
negative exponents belong to irrelevant directions in the phase
space: perturbation components in these directions die out
exponentially. Pure gauge fields on the lattice show a characteristic
Lyapunov spectrum consisting of one third of each kind of
exponents \cite{BOOK}.
This fact reflects the elimination of
longitudinal degrees of freedom of the gauge bosons.
Assuming this general structure of the Lyapunov spectrum we
investigate presently its magnitude only, namely the maximal
value of the Lyapunov exponent, $L_{{\rm max}}$.

The general definition of the Lyapunov exponent is based on a
distance measure $d(t)$ in phase space,
\be
L := \lim_{t\ra\infty} \lim_{d(0)\ra 0}
\frac{1}{t} \ln \frac{d(t)}{d(0)}.
\ee
In case of conservative dynamics the sum of all Lyapunov exponents
is zero according to Liouville's theorem,
\be
\sum L_i = 0.
\ee
We utilize the gauge invariant distance measure consisting of
the local differences of energy densities between two field configurations
on the lattice:
\be
d : = \frac{1}{N_P} \sum_P\nolimits \, \left| {\rm tr} U_P - {\rm tr} U'_P \right|.
\ee
Here the symbol $\sum_P$ stands for the sum over all $N_P$ plaquettes,
so this distance is bound in the interval $(0,2N)$ for the group
SU(N). $U_P$ and $U'_P$ are the familiar plaquette variables, constructed from
the basic link variables $U_{x,i}$,
\be
U_{x,i} = \exp \left( aA_{x,i}^cT^c \right),
\ee
located on lattice links pointing from the position $x=(x_1,x_2,x_3)$ to
$x+ae_i$. The generators of the group are
$T^c = -ig\tau^c/2$ with $\tau^c$ being the Pauli matrices
in case of SU(2) and $A_{x,i}^c$ is the vector potential.
The elementary plaquette variable is constructed for a plaquette with a
corner at $x$ and lying in the $ij$-plane as
\be
U_{x,ij} = U_{x,i} U_{x+i,j} U^{\dag}_{x+j,i} U^{\dag}_{x,j}.
\ee
It is related to the magnetic field strength $B_{x,k}^c$:
\be
U_{x,ij} = \exp \left( \vep_{ijk} a B_{x,k}^c T^c \right).
\ee
The electric field strength $E_{x,i}^c$ is related to the canonically conjugate
momentum $P_{x,i} = \dot{U}_{x,i}$ via
\be
E^c_{x,i} = \frac{2a}{g^3} {\rm tr} \left( T^c \dot{U}_{x,i} U^{\dag}_{x,i} \right).
\ee

\section{Initial states prepared by quantum Monte Carlo }

The Hamiltonian of the lattice gauge field system can be casted into
the form
\be
H = \sum \left[ \frac{1}{2} \langle P, P \rangle \, + \,
 1 - \frac{1}{4} \langle U, V \rangle \right].
\ee
Here the scalar product between group elements stands for
$\langle A, B \rangle = \frac{1}{2} {\rm tr} (A B^{\dag} )$.
The staple variable $V$ is a sum of triple products of elementary
link variables closing a plaquette with the chosen link $U$.
This way the Hamiltonian is formally written as a sum over link
contributions and $V$ plays the role of the classical force
acting on the link variable $U$. The naive equations of motion
following from this Hamiltonian, however, have to be completed
in order to fulfill the constraints
\ba
\langle U, U \rangle &=& 1, \NL
\langle P, U \rangle &=& 0.
\ea
The following finite time step recursion
formula:
\ba
U' &=& U + dt ( P' - \vep U ), \NL
P' &=& P + dt ( V - \mu U + \vep P' ),
\ea
with the Lagrange multipliers
\ba
\vep &=& \langle U, P' \rangle, \NL
\mu &=& \langle U, V \rangle + \langle P', P' \rangle,
\ea
conserves the Noether charge belonging to the Gauss law,
\be
\Gamma = \sum_+ PU^{\dag} - \sum_- U^{\dag}P.
\ee
Here the sums indicated by $+$ run over links starting from,
and those by $-$ ending at a given
site $x$, where the Noether charge $\Gamma$ is defined.
The above algorithm is written in an implicit form, but it can be
casted into explicit steps, so no iteration is necessary \cite{ALGO.IJMP}.

Initial conditions chosen randomly with a given average magnetic energy
per plaquette have been investigated in past years. In the SU(2) case, a
linear scaling of the maximal Lyapunov exponent with the total energy of
the system has been established for different lattice sizes and coupling
strengths \cite{BOOK}.
In the present study we prepare the initial field configurations
from a standard four dimensional Euclidean Monte Carlo program on
a $12^3\times 4$ lattice varying the inverse gauge coupling $\beta$ \cite{SU2}.

We relate such four dimensional Euclidean
lattice field configurations to Minkow\-skian momenta and fields
for the three dimensional Hamiltonian simulation
by the following approach:

First we fix a time slice of the four dimensional lattice.
We denote the link variables in the three dimensional sub-lattice
by $U' = U_i(x,t).$ Then we build triple products on attached handles
in the positive time direction, \newline
\hbox{$U'' = U_4(x,t)U_i(x,t+a)U^{\dag}_4(x+a,t)$.}
We obtain the canonical variables of the Hamiltonian system
by using
\ba
P &=& (U'' - U') / dt, \NL
U &\propto & (U'' + U').
\ea
Finally $U$ is normalized to $\langle U, U \rangle = 1$.

This definition constructs the momenta according to a
simple definition of the timelike covariant derivative.
The multiplication with the link variables
in time direction can also be viewed as a gauge transformation to
$U_4(x,t)=1$, i.e. $A_0=0$ Hamiltonian gauge.

\section{Chaos and confinement  }

\begin{figure}[t]
\centerline{{\psfig{figure=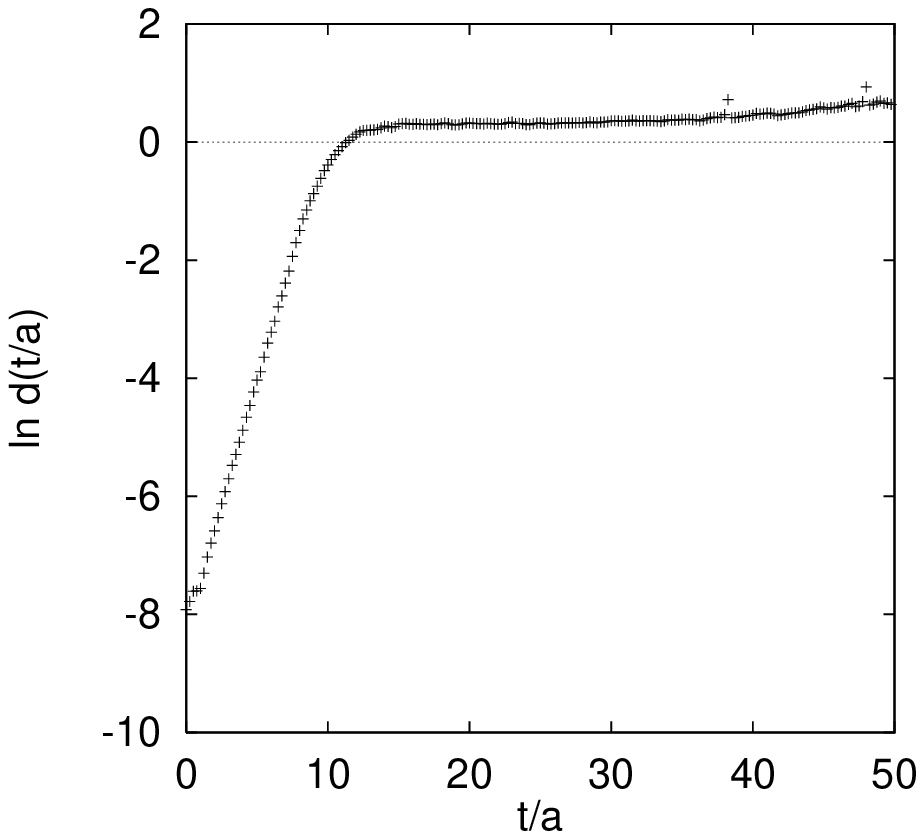,width=5cm}}}
\centerline{{\psfig{figure=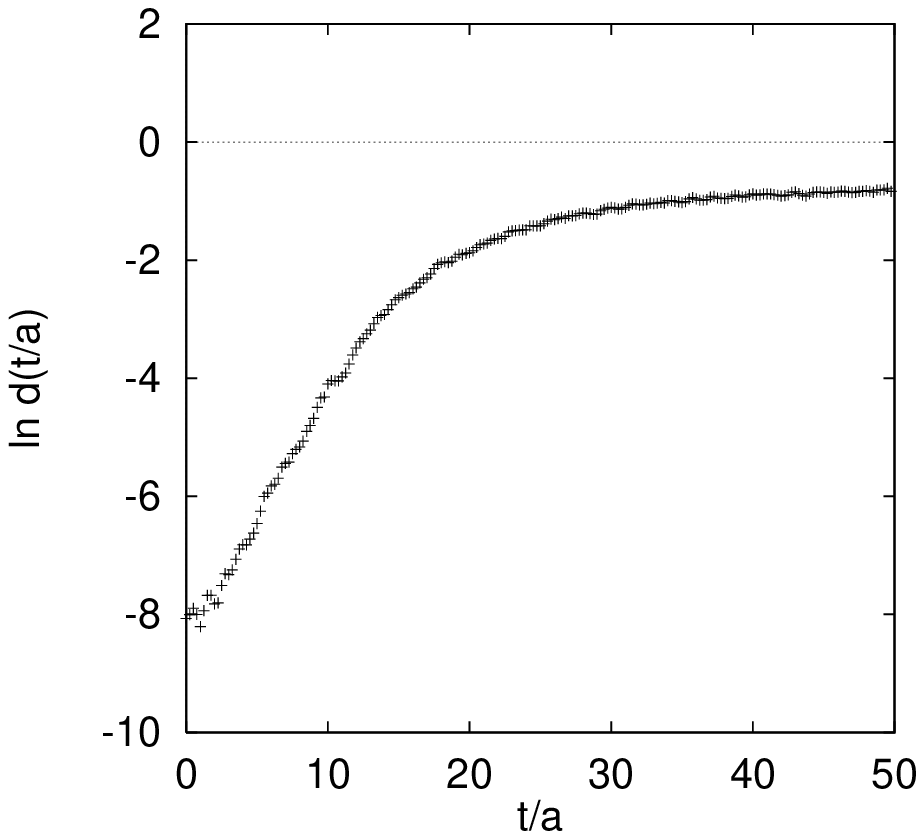,width=5cm}}}
\vspace*{-1cm}
\caption[Fig1]{
  Exponentially diverging distance of initially adjacent U(1) field
  configurations on a $12^3$ lattice prepared at $\beta=0.9$ in the
  confinement phase (top) and at $\beta=1.1$ in the Coulomb
  phase (bottom).
\vspace*{-0.5cm}
\label{Fig1}
 }
\end{figure}

We start the presentation of our results with a characteristic example
of the time evolution of the distance between initially adjacent
configurations. An initial state prepared by a standard four dimensional
Monte Carlo simulation is evolved according to the classical Hamiltonian dynamics
in real time. Afterwards this initial state is rotated locally by
group elements which are chosen randomly near to the unity.
The time evolution of this slightly rotated configuration is then
pursued and finally the distance between these two evolutions
is calculated at the corresponding times.
A typical exponential rise of this distance followed by a saturation
can be inspected in Fig.~\ref{Fig1} from an example of U(1) gauge theory
in the confinement phase and in the Coulomb phase.
While the saturation is an artifact of
the compact distance measure of the lattice, the exponential rise
(the linear rise of the logarithm)
can be used for the determination of the leading Lyapunov exponent.
The naive determination and more sophisticated rescaling methods lead to
the same result.

\begin{figure}
\centerline{\psfig{figure=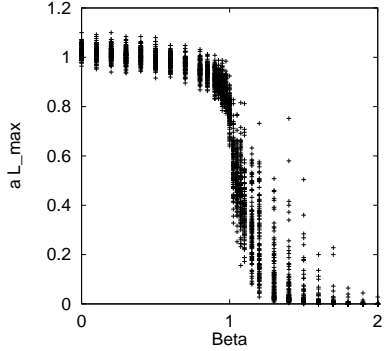,width=5cm}}
\vspace{-0.5cm}
\caption[Fig2]{
   Transition of the leading Lyapunov exponents from 100 U(1) configurations
   as a function of the inverse coupling strength $\beta$.
\label{Fig2}}
\end{figure}

\begin{figure}
\centerline{\psfig{figure=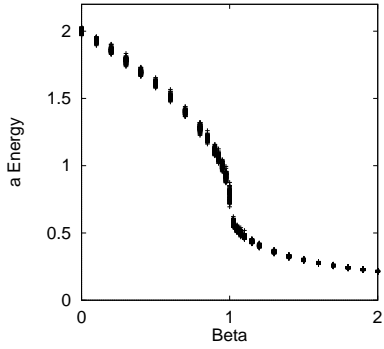,width=5cm}}
\vspace{-0.5cm}
\caption[Fig3]{
   Transition of the plaquette energy as in Fig. 2.
\label{Fig3}}
\end{figure}

\begin{figure}
\centerline{\psfig{figure=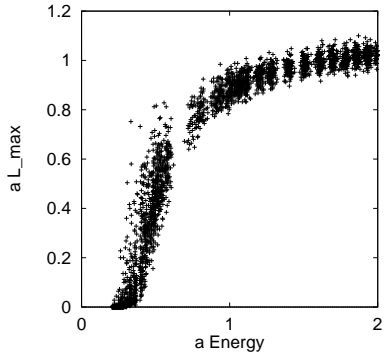,width=5cm}}
\vspace{-0.5cm}
\caption[Fig4]{
   Scatter plots of Lyapunov exponents and plaquette energies for 100 U(1)
   configurations.
\label{Fig4}}
\end{figure}

The main result of the present study is the dependence of the leading
Lyapunov exponent $L_{{\rm max}}$ on the inverse coupling strength $\beta$,
displayed in Fig.~\ref{Fig2} for a statistics of 100 independent U(1)
configurations.  As expected the strong coupling phase, where confinement
of static sources has been established many years ago by proving the area law
behavior for large Wilson loops, is more chaotic.  The transition reflects
the critical coupling to the Coulomb phase. Furthermore the maximal Lyapunov
exponent scatters more pronounced than the average energy per plaquette.
Fig.~\ref{Fig3} shows the somewhat smoother transition of the energy per
plaquette as a function of the inverse coupling strength. Fig.~\ref{Fig4}
depicts the correlation of the Lyapunov exponents and the plaquette energies
for 100 U(1) configurations. The blank area is indicative of the transition
point being presumable of first order.

\begin{figure*}[h]
\centerline{\psfig{figure=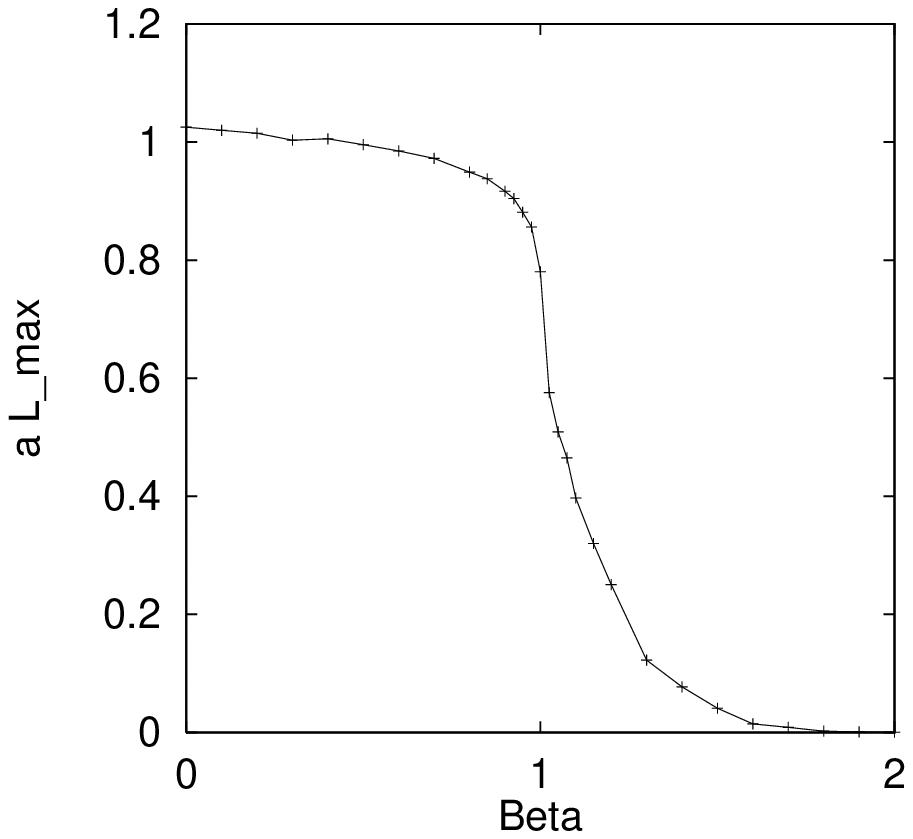,width=5cm}\hspace*{5mm}
\psfig{figure=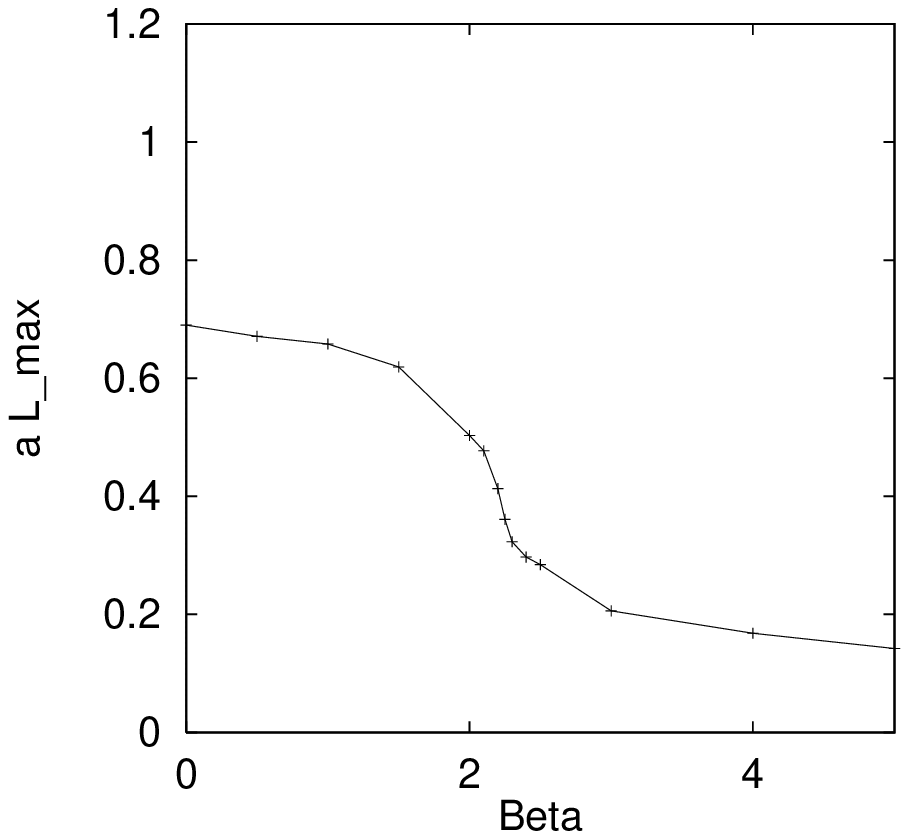,width=5cm}}
\vspace{-0.5cm}
\caption[FIG5]{
Comparison of the average maximal Lyapunov exponent in U(1) gauge theory
with $\beta = 1/g^2$ (left) and in SU(2) gauge theory with $\beta = 4/g^2$
(right) when crossing from the strong to the weak coupling phase.
\label{Fig5} }
\end{figure*}

\begin{figure*}[h]
\centerline{\psfig{figure=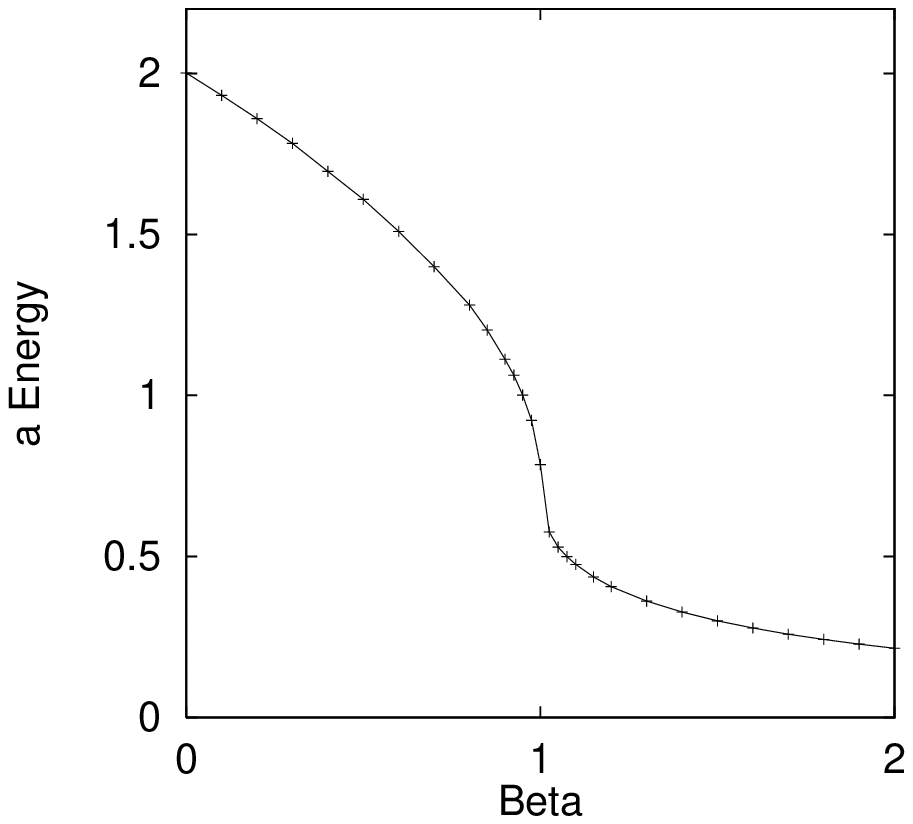,width=5cm}\hspace*{5mm}
\psfig{figure=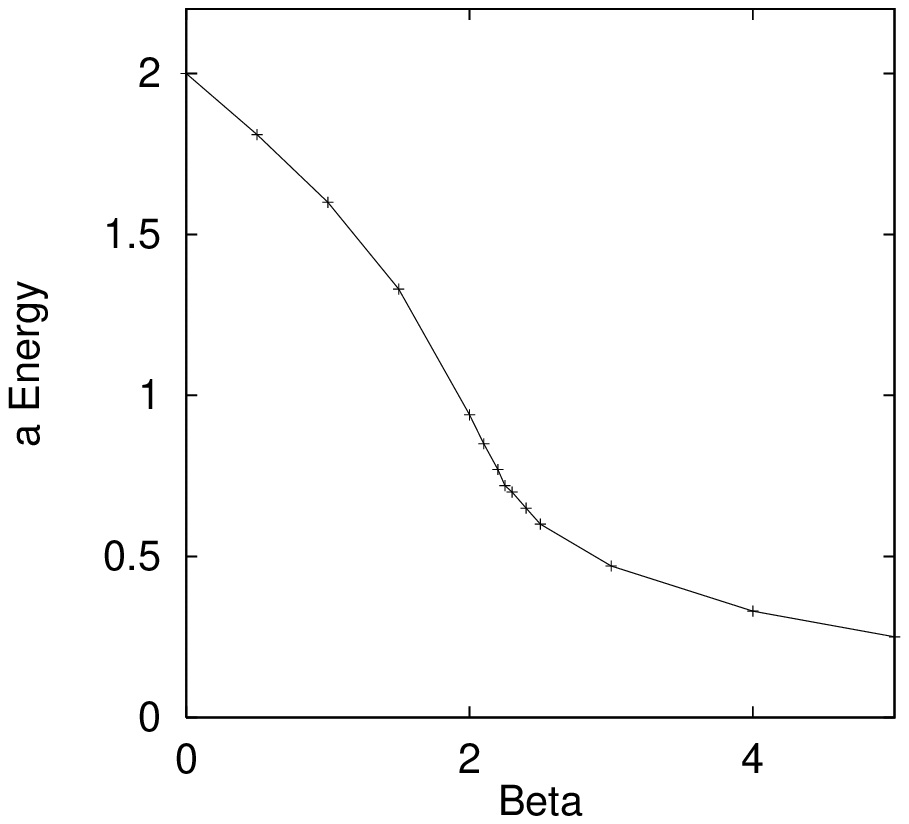,width=5cm}}
\vspace{-0.5cm}
\caption[Fig6]{Comparison of the average plaquette energy as in Fig. 5.
\label{Fig6}}
\end{figure*}

We now turn to a comparison of expectation values of U(1) and SU(2) theory.
Fig.~\ref{Fig5} exhibits the averaged leading Lyapunov exponent between the
strong and the weak coupling regime. The smoother fall-off of the SU(2)
Lyapunov exponent reflects the second order of the finite temperature
transition to a Debye screened phase of free quarks. Fig.~\ref{Fig6}
compares the averaged plaquette energies of both gauge theories
signaling the different order of their phase transitions.

\begin{figure*}[h]
\centerline{\psfig{figure=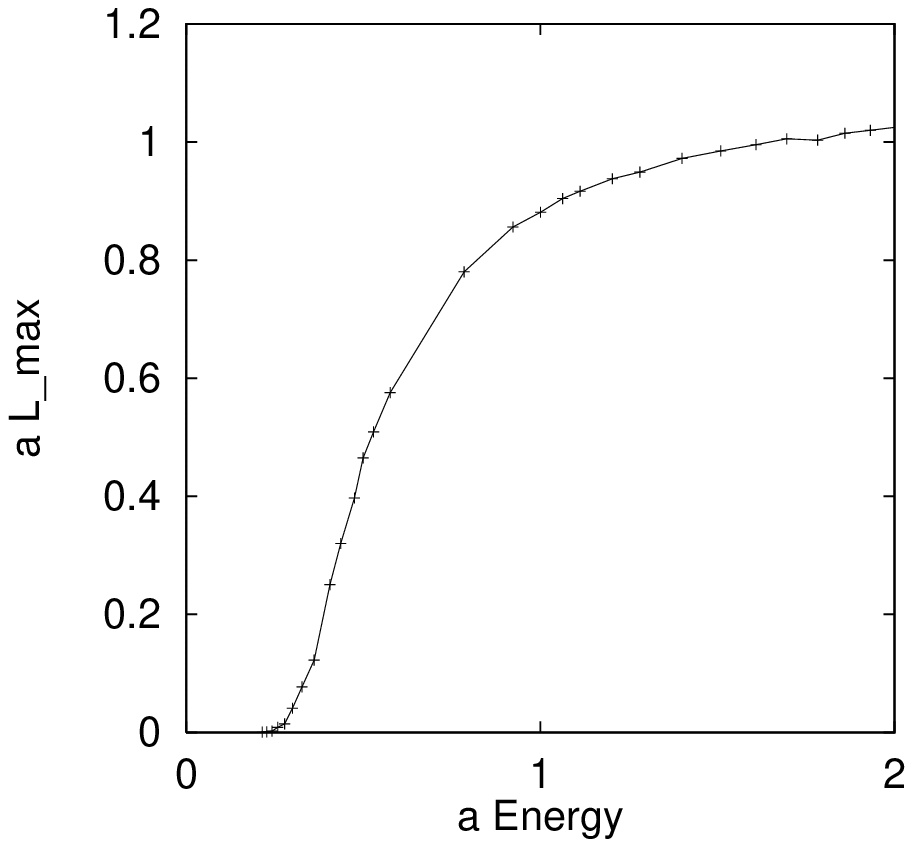,width=5cm}\hspace*{5mm}
\psfig{figure=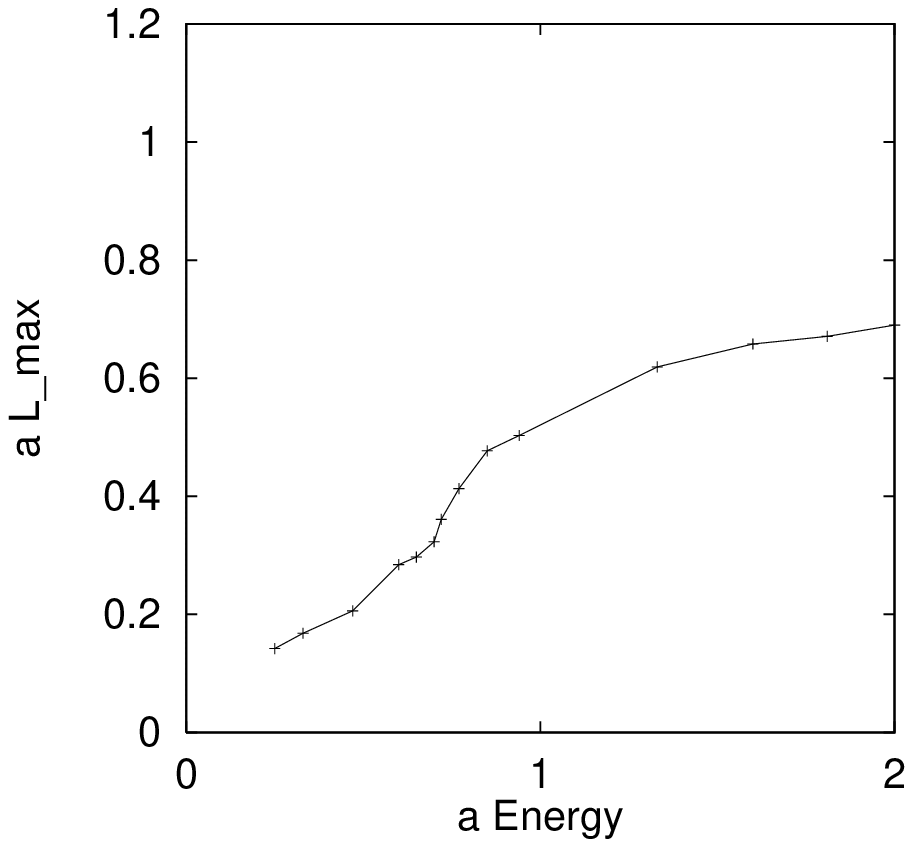,width=5cm}}
\vspace{-0.5cm}
\caption[Fig7]{
  Comparison of average maximal Lyapunov exponents as a function of the
  scaled average energy per plaquette $ag^2E$. The U(1) theory (left)
  shows an approximately quadratic behavior in the weak coupling regime
  whereas the SU(2) theory (right) is approximately linear.
\label{Fig7}}
\end{figure*}

Fig.~\ref{Fig7} shows the energy dependence of the Lyapunov exponents for
both theories.  One observes an approximately linear relation for the SU(2) case while a
quadratic relation is suggested for the U(1) theory in the weak coupling regime.
From scaling arguments one expects a functional relationship between
the Lyapunov exponent and the energy \cite{BOOK,SCALING}
\begin{equation}
L(a) \propto a^{k-1} E^{k}(a) ,
\label{scaling}
\end{equation}
with the exponent $k$ being crucial for the continuum limit of the
classical field theory. A value of $k < 1$ leads to a
divergent Lyapunov exponent, while $k > 1$ yields a vanishing $L$ in
the continuum. The case $k = 1$ is special leading to a finite non-zero
Lyapunov exponent. Our analysis of the scaling relation (\ref{scaling})
gives evidence, that the classical compact U(1) lattice gauge theory
has $k \approx 2$ and with $L(a) \to 0$ a regular continuum theory. The
non-Abelian SU(2) lattice gauge theory signals $k \approx 1$ and stays chaotic
approaching the continuum.

Summarizing we investigated the classical chaotic dynamics of
U(1) and SU(2) lattice gauge field configurations prepared by
quantum Monte Carlo simulation. The maximal Lyapunov exponent
shows a pronounced transition as a function of the coupling strength.
Both for QED and QCD we find that configurations in the strong coupling
phase are substantially more chaotic than in the weak coupling regime.
Our results demonstrate that chaos is present when particles are confined,
but it persists partly also into the Coulomb and quark-gluon-plasma phase.
So far the situation for the gauge fields. An independent analysis of the
fermion fields yields compatible results with respect to quantum chaos 
\cite{QUANTCHAOS}.
\vspace{-2mm}
\section*{ Acknowledgments  }
\vspace{-1mm}
This work has been supported by the Hungarian National Scientific Fund under
the project OTKA T019700 as well as by the Joint American Hungarian Scientific
Fund TeT MAKA 649.

\end{document}